\documentclass[a4paper,11pt]{article}
\pdfoutput=1 

\usepackage{jinstpub} 

\usepackage{lineno}
\usepackage{siunitx}
\usepackage{textcomp}
\usepackage{subcaption}
\DeclareSIUnit\bar{bar}
\usepackage{multirow}

\title{Non-magnetic glass-fiber cryostat for MADMAX prototype tests}

\abstract{MADMAX, an axion dark matter search experiment, is currently in the prototype testing phase. Its working principle is based on the conversion of axions in a magnetic field into photons. This signal is then enhanced by a booster made of dielectric disks placed in front of a mirror. In order to test MADMAX prototypes at cryogenic temperatures in a magnetic field parallel to the disks, a new G-10 glass-fiber cryostat of 0.06 m$^3$ inner volume was designed, tested and used in a CERN magnet. The design allows to minimise cost as well as cooling down and warming up times. Using continuous circulation flow of gaseous helium, the MADMAX prototype was cooled down reproducibly to temperatures below 10 K for more than 24 hours. This procedure allowed, for the first time, to perform a calibration of the booster response and to run a dark matter axion search in a magnetic field at low temperatures. This novel type of cryostat, with a low manufacturing cost, fast and easy to operate, could be used for other purposes beyond MADMAX.}

\keywords{Dark Matter detectors, Cryogenics, Detector design and construction technologies and materials}

\author[a,1]{D.~Kreikemeyer-Lorenzo,\note{Corresponding author.}}
\author[b]{T.~Koettig,}
\author[b]{P.~Borges de Sousa,}
\author[a]{C.~Gooch,}
\author[a]{D.~Kittlinger,}
\author[a]{B.~Majorovits}
\author[a]{J.P.A.~Maldonado,}
\author[c]{P.~Pralavorio,}

\affiliation[a]{Max-Planck-Institut für Physik,\\Boltzmannstr. 8, Garching, Germany}
\affiliation[b]{CERN, Geneva, Switzerland}
\affiliation[c]{Aix Marseille Univ, CNRS/IN2P3, CPPM, Marseille, France}

\emailAdd{dkreike@mpp.mpg.de}

\begin{document}
\maketitle
\flushbottom

\section{Introduction}

The MAgnetized Disk and Mirror Axion eXperiment (MADMAX)~\cite{brun2019} is a future experiment to search for dark matter axions. It is presently in a prototyping phase to validate the new concept of a dielectric haloscope~\cite{caldwell2017} consisting of several parallel dielectric disks and a mirror (the so-called booster). Inside a strong magnetic field ($\gg\SI{1}{\tesla}$), axions would convert into photons and the booster would enhance this photon signal. The booster needs to operate at a low temperature ($<\SI{10}{\kelvin}$) to reduce the contribution of thermal noise. Moreover, low temperatures improve the resonant response of the booster and they reduce signal losses. 

To verify different aspects and components of the dielectric haloscope concept under realistic experimental conditions, a dedicated cryostat was developed to operate various prototype boosters inside a magnetic field and at cryogenic temperatures. To ensure this, a stable pressure and stable temperature below 10 \si{\kelvin} has to be provided for at least one day for a smooth, reliable and reproducible data taking. The stability refers to the absence of sudden variations over the time scale of the data acquisition, as such fluctuations could complicate the analysis of the data. Short cooldown and warm-up times for rapid testing, in the order of hours, are also desired. Moreover, a simple, readily-available cryostat, with low-manufacturing cost, was desired. Since the MADMAX prototypes require the dielectric disks to be aligned parallel to the magnetic field within a dipole magnet, a horizontally oriented cryostat was needed. This differs from most commercial cryostats, which typically have a vertical orientation and lack the necessary diameter to accommodate MADMAX prototype setups. This final constraint stems from the size of the MADMAX prototypes, the largest of which has an outer diameter of 300 mm. All these considerations led to the design of the novel glass-fiber (G-10) cryostat presented here. The G-10 material is diamagnetic, thus allowing to use the cryostat in a strong magnetic field. The cryostat components were delivered in several pieces, which were assembled in the Cryogenics Laboratory at CERN. As cooling source, instead of a more-traditional helium bath, 4.2 \si{\kelvin} helium gas (from a liquid helium transport dewar) in a continuous gas flow was used. This enables a smooth cooldown, reaching a stable sample temperature in a relatively short time.

\section{Cryostat design and experimental setup}

\begin{figure}
    \centering
    \includegraphics[width=\textwidth]{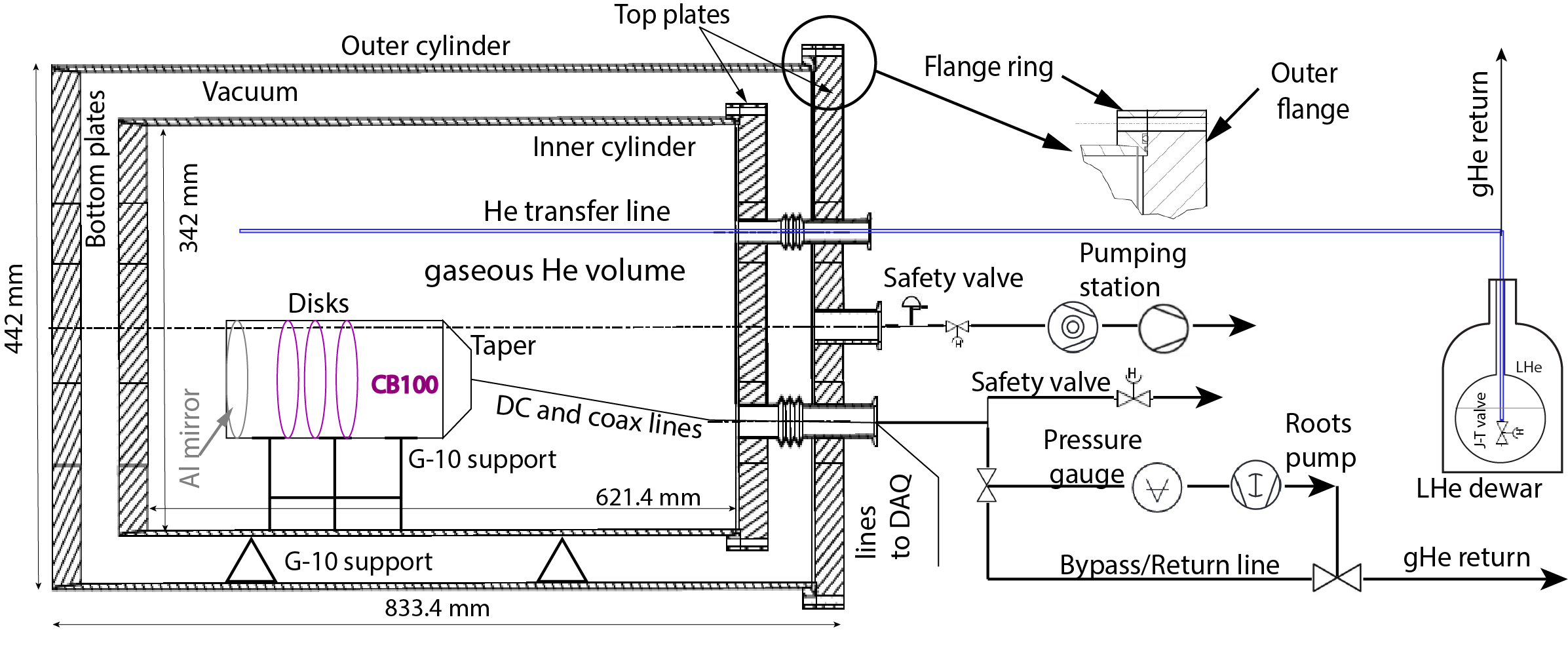}
    \caption{Schematic of the G-10 cryostat. A simplified sketch of a MADMAX prototype is shown inside of the cryostat, with a 100 mm diameter aluminium mirror at the back of the detector and three sapphire disks (in purple). The thickness of both G-10 cylinder walls is 5 mm. The total length of the cryostat is 833.4 \si{\mm} (from outer flange to the bottom of the outer vessel). The inner vessel has an inner diameter of 342 \si{\mm} and the length inside the inner volume is approx. 621 \si{\mm}. }
    \label{fig:cryostat_model}
\end{figure}

\begin{figure}
    \centering
    \includegraphics[width=\textwidth]{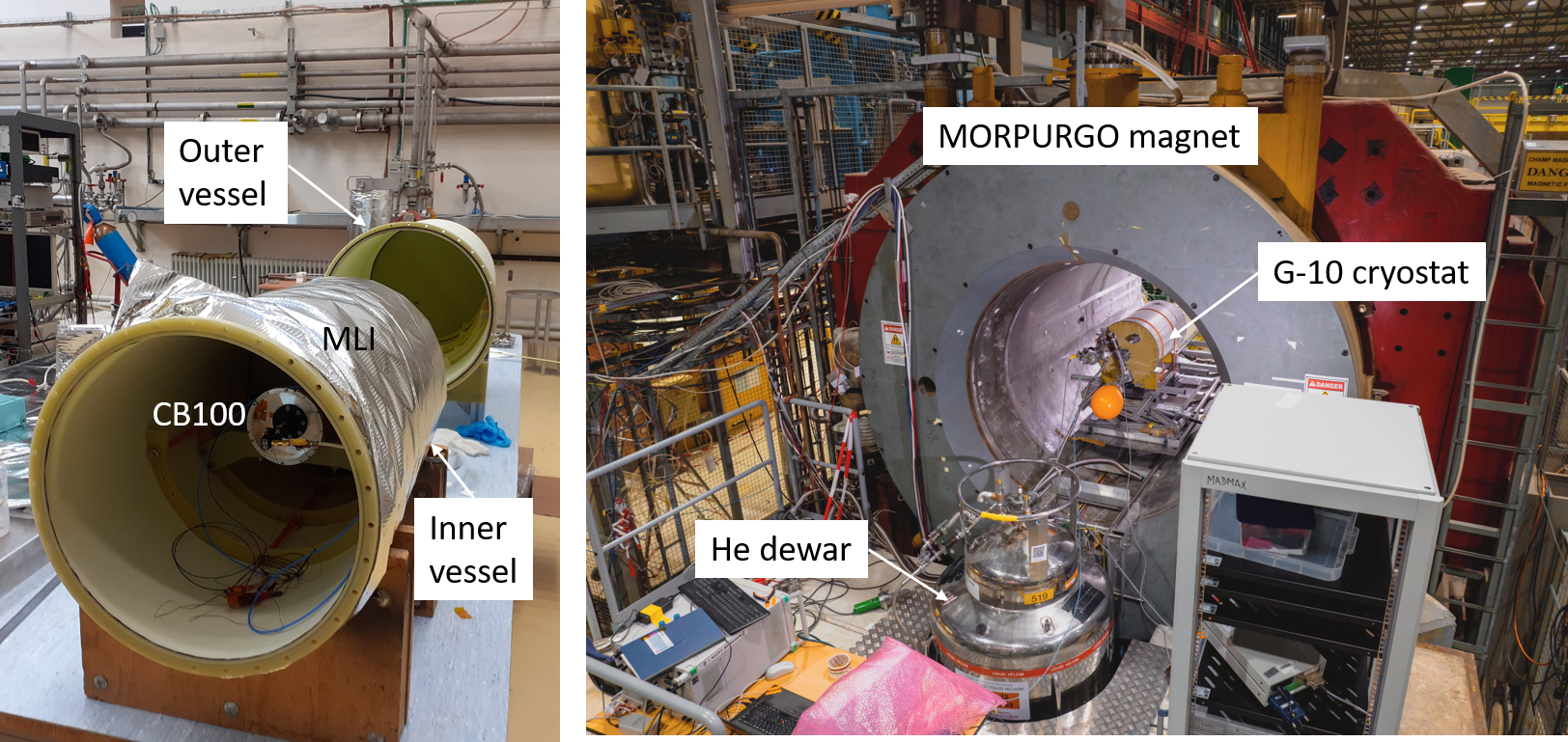}
    \caption{Left: Inner (front) and outer (back) vessels of the G-10 cryostat. The flange ring (glued to the inner cylinder) with 32 M6 holes, can be seen on the front. Inside of the inner vessel, CB100 is already installed. Two MLI blankets are wrapped around the inner vessel. Right: G-10 cryostat inside the 1.6 m warm bore of the CERN MORPURGO magnet during warm-up. The helium dewar can be seen in front of the warm bore aperture, outside of the magnetic field. The transfer line (not shown) was already disconnected.}
    \label{fig:fotosG10}
\end{figure}

The novel cryostat design, shown in Fig.~\ref{fig:cryostat_model}, is based on two glass epoxy laminate material (G-10) vessels separated by an insulation vacuum. Both vessels consist of four parts (manufactured by the company RESARM Engineering Plastics\footnote{RESARM Engineering Plastics SA, Rue Pr\'es-champs 21B, 4671 Barchon, Belgium , www.resarm.com}): cylinder, bottom plate, flange-ring and flange. The three former are glued together with Stycast 2850FT CAT 9. Both bottom plates and  both flanges have a thickness of 30 mm. To minimise heat transfer to the outer cylinder, the inner cylinder is resting on a semi-circular G-10 support structure, with only four small contact points. Two blankets of 10 layers each of multi-layer superinsulation (MLI) are placed around the inner cylinder, as shown in Fig.~\ref{fig:fotosG10} left. The inner and outer walls of both cylinders were coated with Stycast 1266A to minimise possible leaks of helium at room temperature through the porous G-10 material. The top-flanges of both cylinders are joined by two mechanical feedthroughs. One of the feedthroughs is used for the He transfer line, the second one is used for the helium gas return line, as well as to route all the electrical connections (temperature sensors, heater, coax RF cables and amplifiers). The inner vessel is sealed with Indium wire, while the outer vessel is sealed using a rubber gasket. In this simple design, no thermal shield is used.

The G-10 cryostat is designed to be operated in a temperature range between 4 and 10~\si{\kelvin}. The outer vessel (insulation vacuum) is designed to be operated at pressures better than 1x$\SI{e-4}{\milli\bar}$ at room temperature. The insulation vacuum is achieved using a pumping station (with roughing and turbomolecular pumps) attached to a KF25 flange in the middle of the outer flange. A pressure safety disc placed between the cryostat and the pumping station allowed a maximum operating overpressure of 0.4~\si{\bar}. The maximum allowed (absolute) pressure inside the inner vessel is 1.4~\si{\bar}, which is limited by a safety disc similar to the one used for the outer vessel. Prior to every helium transfer, the inner volume of the cryostat was pumped down and purged with high purity helium gas.

The large inner volume inside the cryostat allows to test different setups. In particular, the G-10 cryostat was designed to test the smallest of MADMAX's prototype, Closed Booster 100 (CB100), in a magnetic field at cryogenic temperatures. CB100 consists of a 260 mm long cylinder containing an aluminium mirror and three sapphire disks of 100 mm diameter. A small aperture in CB100 connects the inner volume of the cryostat with the inner volume of the prototype. This was necessary for safety reasons and to ensure better cooling of the sapphire disks via exchange gas. Although the inner volume of the G-10 cryostat is rather large, the transfer line in the upper half of the cryostat needs some space. This limitation has to be integrated in the design. 

For cooling down the inner volume of the cryostat, gaseous helium is used. The helium gas is supplied from a 500 \si{\liter} liquid-helium dewar via a low-loss transfer line with a needle-valve for mass flow control. An integrated Joule-Thomson (JT) valve in the transfer line supply side enables the expansion of the helium before entering into the inner cryostat volume. The decrease of the gas pressure through the JT valve results in the cooling of the gas. The cooling of the cryostat is performed in two steps. For the first step, the liquid helium dewar is pressurised to 1.4 bar (absolute pressure) while the pressure inside of the inner vessel of the cryostat is kept at approximately 1 \si{\bar}. This pressure difference allows for a continuous He flow rate, achieving final temperatures inside the inner vessel close to 50 \si{\kelvin}. For the final step of the cooldown, the pressure in the inner vessel is lowered to values between 200 and 115 \si{\milli\bar} using a roots pump (Pfeiffer\footnote{Pfeiffer Vacuum GmbH, Berliner Strasse 43, 35614 Asslar, Germany, www.pfeiffer-vacuum.com} ACP90) and an exhaust throttle valve (MKS\footnote{MKS Instruments, Inc., 2 Tech Drive, suite 201, Andover, MA 01810, USA, www.mks.com} 653B) to maintain a stable pressure. This now larger gradient in pressure allows to reach temperatures below 10 \si{\kelvin}. In order to monitor the temperature evolution, three magnetic-field compatible CERNOX temperature sensors (model CX-1050) from LakeShore\footnote{LakeShore Cryotronics, 550 Tressler Dr, Westerville, OH 43082-7587, USA , www.lakeshore.com} were placed on different parts inside the cryostat and on different setup components during the seven cooldowns carried out with this cryostat. 

For the final measurements with CB100 inside the magnetic field, the G-10 cryostat was placed in the 1.6~T magnetic field of the MORPURGO dipole magnet at CERN~\cite{MORPURGO1979411}. In both locations (Cryogenics Laboratory and MORPURGO magnet), the helium gas was pumped out directly into the CERN He recovery system. A picture of the setup inside the large warm bore of 1.6 m diameter aperture of the magnet is shown in Fig.~\ref{fig:fotosG10} right.  The G-10 cryostat is placed 1 m inside the magnet, allowing the CB100 booster (in the back of the cryostat) to reach the central section of the magnet, where the magnetic field is strongest. 

The heat load corresponding to thermal radiation of the outer vessel onto the inner vessel, taking into account the 20 layers of MLI used in the setup, is estimated to be 8~\si{\watt}. Heat loads due to cabling and the support structure of the inner vessel are estimated to be less than 1~\si{\watt}. 

\section{Commissioning and Results}

A summary of the different cooldowns performed with the G-10 cryostat is shown in Table~\ref{tab:table1}. A total of seven cooldowns were realised.  Three commissioning cooldowns were conducted to find optimal parameters to perform the cooling. After that, three additional cooldowns were accomplished to test different components and calibrations steps relevant for the axion dark matter search. Finally, the equipment was transported from the Cryogenics Laboratory at CERN, where assembly and cooldowns 1-6 were performed, to the MORPURGO magnet, where a final cooldown was carried out inside the magnet. During this run, a first dark matter search using CB100 was realized after reaching the end temperature. For this last experimental run, the tubing between the cryostat and the pump, as well as the helium return line, were extended for space reasons from 2 m to 4 m. This small change in the setup configuration resulted in a different pressure drop in the cryostat, which led to a temperature increase of 2~\si{\kelvin} with respect to previous cooldowns. Additionally, in this new location, the outlet of the roots pump had to work against a slightly higher pressure (20 to 30 \si{\milli\bar}), due to a different He recovery system that in the laboratory. This change might have contributed as well to the higher final temperature.

During cooldowns 1-6, all three sensors showed temperature differences smaller than 0.5 \si{\kelvin} for the different passive elements in the cryostat. For comparison between cooldowns, one sensor was always kept on the aluminium mirror at the rear of CB100. Additionally, the pressure in the insulation vacuum was always better than 5x$\SI{e-4}{\milli\bar}$ , with typical values around 9x$\SI{e-6}{\milli\bar}$ at cold temperatures. 

\begin{table}[ht]
  \begin{center}

    \small
    \begin{tabular}{|c|c|c|c|c|c|} 
      \hline
      \textbf{  } & \textbf{Lowest T (mirror)} & \textbf{Time below 10 K} & \textbf{B-field} & \textbf{Goal} \\
      \hline
      1 & 5.6 \si{\kelvin} & 24 h & 0 \si{\tesla}  & commissioning \\
      \hline
      2 & 7.0 \si{\kelvin} & 7 h & 0 \si{\tesla}  & commissioning \\
      \hline
      3 & 6.6 \si{\kelvin} & 22 h & 0 \si{\tesla}  & commissioning \\
      \hline
      4 & 4.3 \si{\kelvin} & 26 h & 0 \si{\tesla}  & calibration \\
      \hline
      5 & 4.2 \si{\kelvin} & 35 h & 0 \si{\tesla} & calibration \\
      \hline
      6 & 4.2 \si{\kelvin} & 30 h & 0 \si{\tesla} & calibration \\
      \hline
      7 & 6.4 \si{\kelvin} & 29 h & 1.0-1.6 T & axion search \\
      \hline
    \end{tabular}
  \end{center}
    \caption{Summary of commissioning and test cooldowns. The temperatures refer to the temperature sensor attached to the Al mirror of CB100, which represents the general setup temperature. Typical times for cooldown are between 12 and 15 hours and 24 hours for warm-up.}
    \label{tab:table1}
\end{table}

\begin{figure}
    \centering
    \includegraphics[width=\textwidth]{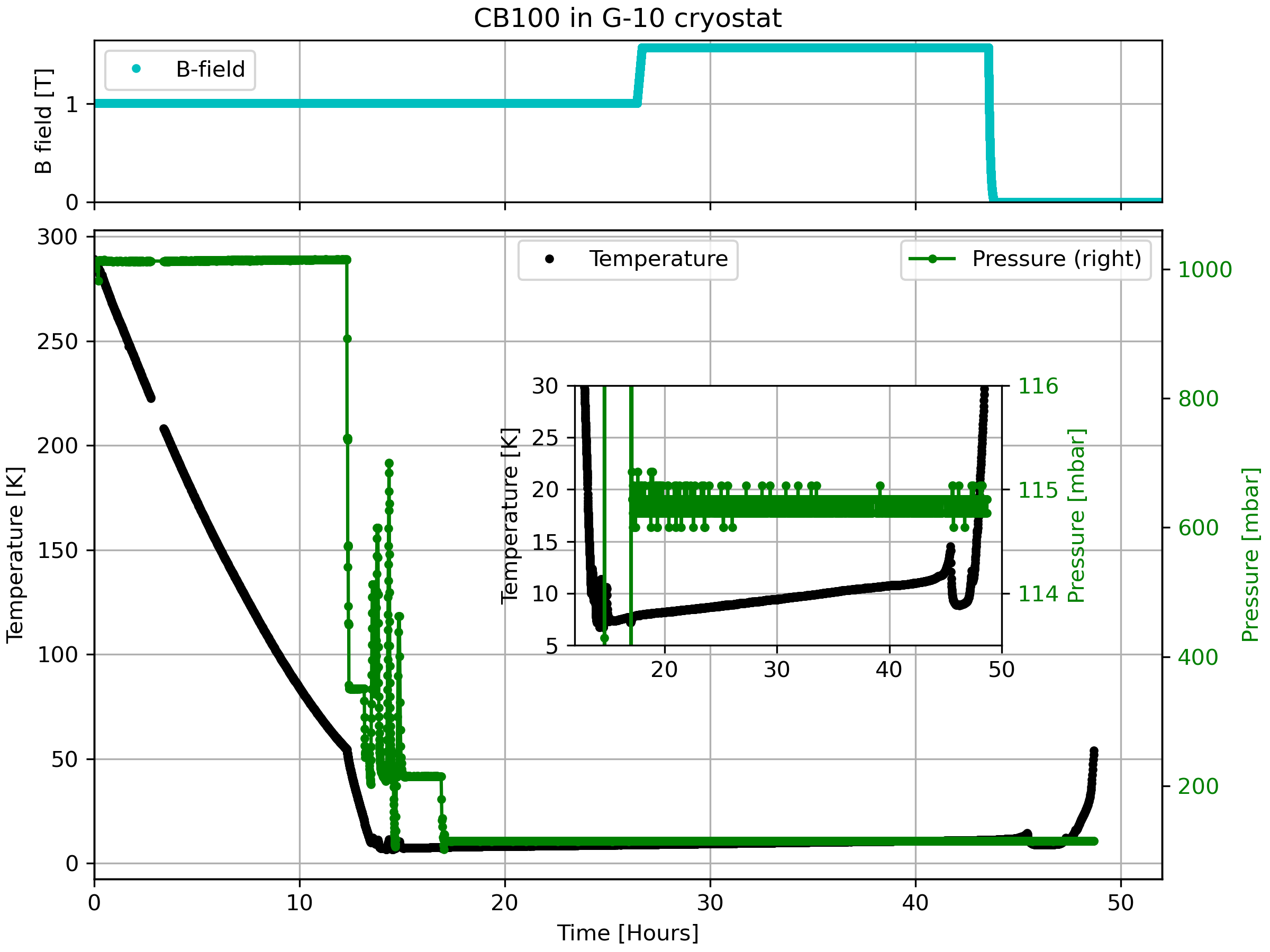}
    \caption{Top: Magnetic field as a function of time. Bottom: Temperature and pressure evolution of the G-10 cryostat during a cooldown inside the magnet. After approx.~12.5 hours, the cryostat reached a temperature close to 50 \si{\kelvin}. To cool down the cryostat to lower temperatures, the pressure difference between the helium dewar, cryostat and return line was increased by pumping on the inner vessel. A final temperature of 6.4 \si{\kelvin} was reached shortly after. The insert shows the lowest temperature range, where it can be seen that the temperature continuously increases, but it stays below 10 \si{\kelvin} for more than 24 hours. The pressure is controlled by a flow-control valve.}
    \label{fig:all_overview}
\end{figure}

Figure~\ref{fig:all_overview} (bottom) shows the temperature and pressure as a function of time during cooldown 7 inside the B-field, during cold operation, and at the beginning of the warm-up phase. After pressurising the helium dewar to 1.4 \si{\bar}, the valve at the transfer line was opened and helium transfer immediately started. After 12.5 hours, the temperature stabilised around 54 \si{\kelvin}. The pressure in the inner vessel was then gradually lowered down to 115~\si{\milli\bar}, as shown in the inset in figure~\ref{fig:all_overview}. After one hour, a final temperature of 6.4 \si{\kelvin} was reached. During the cold operation phase, data taking for axion search was performed. The magnetic field (shown in figure~\ref{fig:all_overview}, top) was initially set to 1~\si{\tesla}. After several hours of stable operation, the magnet was ramped up to a higher current, corresponding to a 1.6 \si{\tesla} magnetic field. While the temperature was still low, the magnet was switched off, in order to take reference data without magnetic field.

The inset in figure~\ref{fig:all_overview} shows the temperature during "cold operation". The gradual rise in temperature was caused by the decreasing liquid helium (LHe) level in the supply dewar. As the static height of the liquid level dropped, it affected the pressure difference in the transfer system. After 45 hours, the temperature increased further when the LHe level fell below the transfer line inlet, resulting in only cold helium vapor being transferred. Just before the full warm-up begins, a sudden drop in temperature can be seen. When the transfer line on the dewar supply side rises above the liquid helium level, a relatively rapid depressurization of the remaining liquid helium occurs, causing the pressure in the dewar to drop from 1.4 \si{\bar} to approximately 1.05 \si{\bar}. This sudden pressure drop induces the liquid helium to boil, adjusting to the lower pressure and corresponding saturation temperature. These flow conditions are partially transferred through the line into the cryostat, leading to this short-term temperature drop.

During the last cooldowns, it was possible to optimise the process to increase the time at low temperatures. However, the maximum time below a temperature of 10 \si{\kelvin} was 35 hours. This time is mostly limited by the size of the helium dewar and in particular, by the amount of liquid helium that can be accessed with the available transfer line. 

After seven cold-cycles, no signs of material ageing has been observed. However, a rise in the vacuum insulation pressure with every new cooldown could be observed, implying that the glass-fiber walls of the inner vessel were saturating with helium. A long pumping time after the last experimental run showed a recovery of the material by allowing helium outgassing at room temperature.

\section{Conclusions}

A novel glass-fiber (G-10) cryostat was designed, commissioned and used in a 1.6 \si{\tesla} magnetic field. Utilising continuous helium gas flow as a cooling method, the cryostat achieved temperatures of approximately 4 \si{\kelvin} during commissioning and calibration tests and 7 \si{\kelvin} during the measurements taken inside the magnet. The payload inside the cryostat remained satisfactorily below 10~\si{\kelvin} for more than 24 hours, enabling to perform a cryogenic axion search using a MADMAX prototype. Below 10 \si{\kelvin}, contributions from thermal noise are low enough for the experiment. Results from this experimental run will be detailed in a forthcoming publication. The "cold-operation" time can be increased further by using a longer helium transfer line into the transport He dewar that would allow to access all of the 500~\si{\liter} of liquid helium. It was also demonstrated in the Cryogenics Laboratory that using a more powerful pump to decrease the pressure in the cryostat further leads to even lower temperatures. Finally, although the G-10 material was expected to degrade and become brittle with each cooling cycle, no signs of material ageing were observed after seven cycles.

An important consideration was the required temperature and pressure stability during data acquisition. Pressure stability was effectively maintained using the throttle-valve. As demonstrated in the previous section, the temperature exhibited a gradual increase, remaining stable within 1\% between 18 hours and 45 hours, with a controlled rise of 0.2 \si{\kelvin}/h. This smooth and predictable variation is not expected to compromise the integrity of the data analysis. 

The G-10 cryostat, designed for magnetic-field compatibility, combines simplicity, fast cooling, and low manufacturing cost, making it ideal for rapid cryogenic testing. However, it is not intended to replace conventional cryostats for applications requiring e.g.~precise temperature control or extended operation times. One additional aspect to be considered is the total cost of the experiment. Although the manufacturing cost is very low, the system's helium consumption (approximately 16~\si{\liter\per\hour} for a heat load of ~9 W) is not negligible and has to be taken into account. Another relevant aspect to mention is the saturation of helium in the cryostat walls. Due to the porosity of the G-10 material, the insulation vacuum was degraded with every cooldown. Long pumping times were needed to recover a good vacuum level.  

In summary, the G-10 cryostat offers a solution to test a wide range of setups (beyond MADMAX) inside or outside of a magnetic field. Using the described technology, it would be possible to go to even larger dimensions, for instance to fit a future planned MADMAX prototype. This larger prototype called Open Booster 300 (OB300) has an outer diameter of 400~\si{\mm} and a length of 600~\si{\mm}. In summary, this work underscores the potential of this novel cryostat design for rapid prototyping and testing in cryogenic research, especially for simple and fast demonstration of applications in a cold environment.

\acknowledgments
The authors would like to thank the CERN magnet team for the support during the measurements at CERN and Agostino Vacca for his invaluable assistance at the Cryogenics Laboratory. We acknowledge the support by the Max Planck Society.



%
\bibliographystyle{hunsrt}
\bibliography{G10}   

\begin{thebibliography}{1}

\bibitem{brun2019}
{MADMAX Coll.}
\newblock \textit{A new experimental approach to probe QCD axion dark matter in the mass range above 40 $\mu$eV}.
\newblock {\em Eur. Phys. J.}, C79(3):186, 2019, \href{https://arxiv.org/abs/1901.07401}{1901.07401}.

\bibitem{caldwell2017}
A.~Caldwell, G.~Dvali, B.~Majorovits, A.~Millar, G.~Raffelt, J.~Redondo, O.~Reimann, F.~Simon, and F.~Steffen.
\newblock \textit{Dielectric Haloscopes: A New Way to Detect Axion Dark Matter}.
\newblock {\em Phys. Rev. Lett.}, 118(9):091801, 2017, \href{https://arxiv.org/abs/1611.05865}{1611.05865}.

\bibitem{MORPURGO1979411}
M.~Morpurgo.
\newblock \textit{A large superconducting dipole cooled by forced circulation of two phase helium}.
\newblock {\em Cryogenics}, 19(7):411--414, 1979.

\end{thebibliography}






\end{document}